\documentclass[prl,preprint,showpacs,superscriptaddress]{revtex4}
\usepackage{epsfig}

\usepackage{amsmath}
\usepackage{dcolumn}
\usepackage{bm}
\usepackage{color,colordvi,graphicx} 

\newcommand{\hide}[1]{}

\newcommand{\eq}[1]{Eq.\,(\ref{#1})}

\newcommand{\fig}[1]{Fig.\,\ref{#1}}

\newcommand{\beq}{\begin{equation}}
\newcommand{\eeq}{\end{equation}}

\begin{document}
\title{Coherent Graphene Devices: Movable Mirrors, Buffers and Memories}

\author{L. Zhao}
\affiliation{Department of Physics, University of Connecticut,
Storrs, Connecticut 06269}

\author{S. F. Yelin}
\affiliation{Department of Physics, University of Connecticut,
Storrs, Connecticut 06269} \affiliation{ITAMP, Harvard-Smithsonian Center
for Astrophysics, Cambridge, Massachusetts 02138}

\date{\today}
\begin{abstract}
We theoretically report that, at a sharp electrostatic step potential in graphene, massless Dirac fermions can obtain Goos-H\"{a}nchen-like shifts under total internal reflection. Based on these results, we study the coherent propagation of the quasiparticles along a sharp graphene \emph{p-n-p} waveguide and derive novel dispersion relations for the guided modes. Consequently, coherent graphene devices (e.g. movable mirrors, buffers and memories) induced only by the electric field effect can be proposed.
\end{abstract}


\maketitle
A totally reflected light beam can laterally shifted from the position expected by a geometrical trajectory, an effect termed the ``Goos-H\"{a}nchen'' (GH) shift \cite{goos}. This effect can be qualitatively attributed to evanescent wave penetration into the medium with a smaller refractive index. Some theories, such as stationary phase models \cite{art}, current-flux models \cite{cfm}, etc., have been developed to formulate this effect.  Recently, the exploration of negative refraction brought on new interest in this subject since it can lead to negative GH shifts \cite{nghs1, nghs2}. Based on negative GH shifts and waveguide theory \cite{wg}, the ``trapped rainbow'' storage of light in a tapered left-handed heterostructure (LHH) offers a novel mechanism to bring light to a complete standstill efficiently and coherently,  which may lead to applications in optical data processing, storage and quantum optical memories \cite{tr}.

Graphene, a monolayer of carbon atoms, has attracted great interest since it was successfully prepared \cite{g01, grev1, grev2}. The interaction between electrons and the two-dimensional (2D) honeycomb lattice of carbon atoms results in a gapless band structure near Dirac cones with a linear (photon-like) electron-hole dispersion, which gives rise to new quasiparticles. Besides, the hopping of electrons between the two triangular sublattices introduces pseudospin 1/2 and chirality to the quasiparticles, which can be analogous to massless neutrinos \cite{grev1, grev2}. Therefore, the quasiparticles can possess the low-energy dynamics effectively described by a Dirac equation with Fermi velocity $v_{F} \approx 10^{6}$ m/s, and they are also called massless Dirac fermions. Furthermore, long phase coherence length \cite{lcl} and tunable Fermi levels based on electric field effects \cite{g01, pn} have been experimentally observed in graphene. Based on the above properties, recent theoretical research has focused on the coherent transport properties of quasiparticles in the electrically gated graphene structures, including Klein tunneling \cite{klein1, klein2}, focusing of electron flow \cite{nr1}, confined states and resonant tunneling in quantum wells \cite{qw1, qw2}, coherent transmission through graphene strips \cite{gstrip1, gstrip2}, etc. All these studies imply that it might be hard to fully and coherently control the group velocity of the massless quasiparticles in the graphene structures induced by electric field effects alone. In some cases, graphene quantum dots \cite{qd1, spin, qd2} or magnetic fields \cite{bf} can be introduced, but for trapping only.

\begin{figure}[ht]
   \centerline{ \includegraphics[clip,width=1\linewidth]{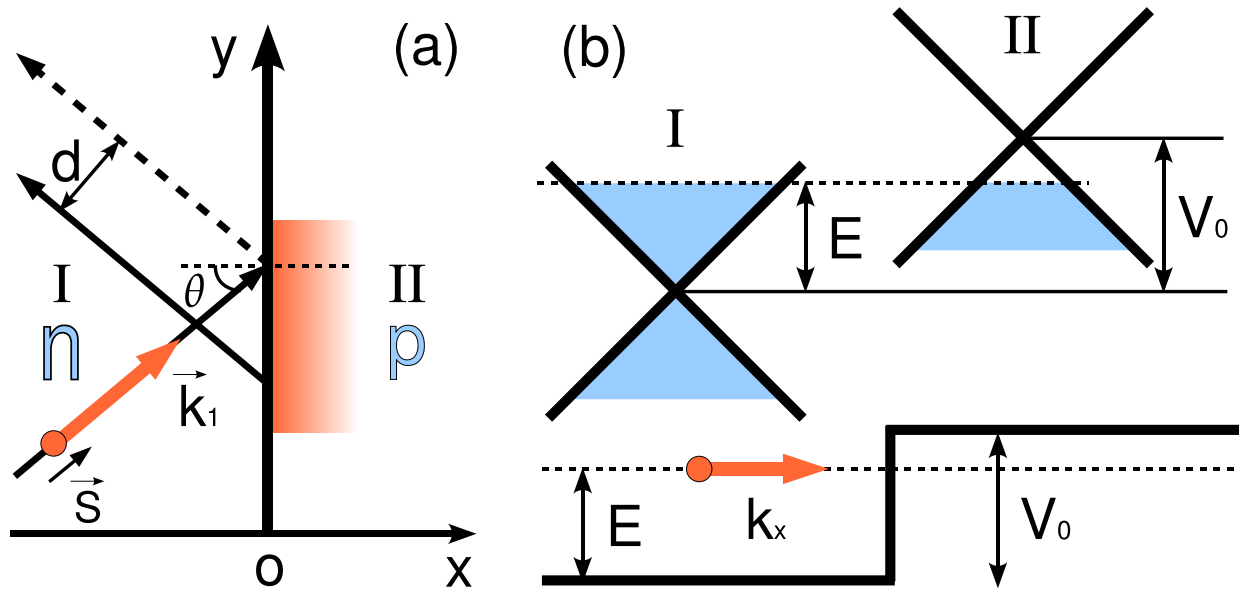}}
         \caption{(Color online) (a) Total internal reflection and negative GH-like 
shift at a sharp \emph{p-n} junction, where $\theta$ is the incident angle and $d$ is the  lateral displacement of the shift. The gradient intensity area represents the evanescent field. On the \emph{n} side, the pseudospin $\boldsymbol{S}$ of an electron-like quasiparticles is parallel to the momentum $\boldsymbol{k_{1}}$. (b) Low-energy linear spectrum and potential diagram.}
   \label{fig:pnj}
\end{figure}

In this paper, however, we have discovered the possibility to coherently decelerate, trap and re-accelerate the massless Dirac fermions along a sharp graphene \emph{p-n-p} waveguide by using a tunable electric field. The waveguide in our scheme can be produced in a bulk graphene sheet by electrostatic gates only, which avoids the complex effects of the graphene strip edges \cite{gstrip1, gstrip2, qd1, spin} and the magnetic fields \cite{bf}. Our results are the electronic counterpart to the trapped rainbow effect in optics \cite{tr}. Before discussing the coherent propagation of quasiparticles along a sharp graphene \emph{p-n-p} waveguide, we first examine the total internal reflection (TIR) of quasiparticles at a sharp electrostatic step potential and focus on the GH-like shift, which is inspired by the similarities between the reflection and refraction of photons and those of graphene quasiparticles at interfaces \cite{nr1, grev2}. 

In our scheme, we theoretically consider the coherent ballistic transport of quasiparticles and ignore the disorder as was also done in Refs. \cite{klein1, klein2, nr1, qw1, qw2,gstrip1, gstrip2, qd1, spin, qd2, bf}. Moreover, since the electron-electron interaction can not severely distort the shapes of constant energy contours of low energy Fermi levels \cite{in}, it can be ignored as well \cite{klein1, klein2, nr1, qw1, qw2,gstrip1, gstrip2}. We also assume that, the edge smearing length of the sharp potential is much smaller than the incident quasiparticle wavelength, but much larger than the lattice constant of graphene, which can prevent intervalley scatterings \cite{klein2}. The quasiparticle dynamics is therefore governed by the 2D Dirac equation
\begin{equation}\label{dirac}
[\hbar v_{F} (\boldsymbol{\sigma} \cdot \boldsymbol{k})+V(x,y)] \psi\;=\; E\psi, 
\end{equation}
where $v_{F} \approx 10^{6}$ m/s is the Fermi velocity, $\boldsymbol{\sigma}=(\sigma_{x},\sigma_{y})$ are the Pauli matrices, $\boldsymbol{k}=(k_{x},k_{y})=-i \boldsymbol{\nabla}$ is the momentum operator, the potential $V(x,y)=V_{0}\textit{H}(x)$ is $y$-independent with the Heaviside step function $\textit{H}$, and $E$ is the eigenenergy. Because a unit cell of the graphene honeycomb lattice contains two sublattices A and B, the state $\psi$ is expressed by the two-component spinor 
$\psi=(\psi_{A}(x,y),\psi_{B}(x,y))^{T}$, where $\psi_{A}(x,y)$ and $\psi_{B}(x,y)$ represent the smooth enveloping functions at each sublattice. The conservation of $k_{y}$ leads to $\psi_{m}(x,y)=\phi_{m}(x)e^{ik_{y}y},\ m=A,\ B$. Thus, \eq{dirac} gives 
\begin{subequations}\label{ab}
\begin{eqnarray}
-i \hbar v_{F} (\partial \phi_{B}/ \partial x+k_{y}\phi_{B})=(E-V_{0}H(x))\phi_{A}, \\
-i \hbar v_{F} (\partial \phi_{A}/ \partial x-k_{y}\phi_{A})=(E-V_{0}H(x))\phi_{B}. 
\end{eqnarray}
\end{subequations}
The wave functions in the two different regions can be written in terms of incident, reflected and evanescent waves with the incident angle $\theta$. Figure \ref{fig:pnj} shows a \emph{p-n} junction for clarification. We define $k_{1}=|E|/\hbar v_{F}$,  $k_{2}=|E-V_{0}|/\hbar v_{F}$, and $k_{y}=k_{1}\sin\theta$. In region $I$, we have
\begin{eqnarray} \label{s1}
&& \psi_{I} = \psi_{I}^{i}+\psi_{I}^{r} \\
&& = \frac{1}{\sqrt{2}}\binom{1}{s e^{i \theta}}e^{i k_{x} x + i k_{y} y}+\frac{r}{\sqrt{2}} \binom{1}{s e^{i (\pi-\theta)}}e^{-i k_{x} x + i k_{y} y} \nonumber 
\end{eqnarray}
where $k_{x}=k_{1}\cos\theta$, $s=\textrm{sgn}(E)$,  $\psi_{1}^{i}$ ($\psi_{1}^{r}$) is the incident (reflected) wave function and $r$ is the reflection amplitude. In region $II$, the evanescent wave is 
\begin{equation}\label{s2}
\psi_{II}= \frac{t}{\sqrt{2}} \binom{1}{i s^\prime (\alpha+k_{y})/ k_{2}}e^{- \alpha x + i k_{y} y}, 
\end{equation}
where the decay constant $\alpha=(k^{2}_{y}-k^{2}_{2})^{1/2}=(k^{2}_{1} \sin^{2} \theta-k^{2}_{2})^{1/2}$, $s^\prime=\textrm{sgn}(E-V_{0})$, and $t$ is the transmission amplitude. Because $\alpha$ and $k_{x}$ are nonzero real numbers, we can obtain the inequalities
\begin{equation}\label{ie}
\frac{(E-V_{0})^{2}}{\hbar^{2} v^{2}_{F}} < k^{2}_{y} < \frac{E^{2}}{\hbar^{2} v^{2}_{F}} \Longrightarrow  0 < V_{0} < 2E,
\end{equation}
which is a necessary condition to obtain evanescent waves. When condition (\ref{ie}) is satisfied, the wavevector $k_{2}$ in region $II$ is smaller than $k_{1}$ in region $I$, which is analogous to the propagation of light from a medium having a larger refractive index into one having a smaller refractive index. Hence, the critical angle $\theta_{c}$ for the TIR can be defined by $\sin \theta_{c}=k_{2}/k_{1}$. One can calculate the reflection amplitude $r$ with the boundary conditions of the wave functions at $x=0$,
\begin{subequations}\label{rt}
\begin{eqnarray}
1+r &=& t,  \\
s \sin \theta_{c} (e^{i \theta}-r e^{-i \theta}) &=& i s^\prime t (\delta + \sin \theta),
\end{eqnarray}
\end{subequations}
where $\delta=\sqrt{ \sin^{2} \theta- \sin^2 \theta_{c}}$. From \eq{rt}, we can get
\begin{equation}\label{r}
r=|r|\exp[i\varphi_{r}]= \frac{ \sin \theta_{c}  \cos \theta + i \gamma}{ \sin \theta_{c}  \cos \theta - i \gamma},
\end{equation}
where $\gamma = \sin \theta_{c}  \sin \theta - s s^\prime (\delta+\sin \theta)$ and $\vert r \vert=1$.

\begin{figure}[ht]
\centerline{ \includegraphics[clip,width=0.7\linewidth]{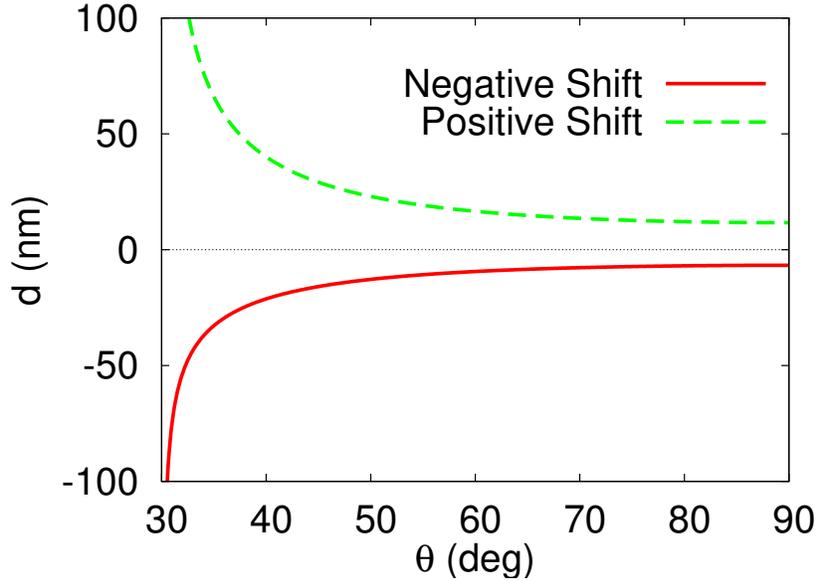}}
  \caption{(Color online) Negative GH-like shift (solid) at the \emph{p-n} junction 
(interband tunneling) with $s=1$ and $s^\prime =-1$. Positive GH-like shift (dashed) at the \emph{n-$n^\prime$} junction (intraband tunneling) with $s= s^\prime =1$. In region $I$, the wavelength of the quasiparticles is $\lambda_{1}=2 \pi /k_{1}=100$ nm. In region $II$, $\lambda_{2}=2 \pi /k_{2}=200$ nm.  The critical angle is $\theta_{c}=30\,^{\circ}$.}
  \label{fig:d}
\end{figure}

Some further discussions follow. First, we compare the TIR of the massless Dirac fermions in graphene and that of the Dirac electrons in high energy physics \cite{de1, de2} because both of them are (pseudo)spin $1/2$ fermions. For the TIR of the Dirac electrons, the incident and reflected momentum states are well defined due to the conservation of momentum, but the spin has two states (e.g. parallel and anti-parallel to the momentum direction), which means there are two eigenstates for incidence and reflection, respectively. Different eigenstates have different phase shifts, which can cause the splitting. However, for the TIR of the massless (neutrino-like) quasiparticles in graphene, the incident and reflected momentum states are well defined, and as a consequence, the pseudospin states are well defined due to the chirality. This means there is only one eigenstate for incidence and reflection, respectively. Similar to Dirac electrons, one can calculate and find the rotation of the pseudospin of the quasiparticles can not produce any extra phase shifts and the reflection amplitude $r$ provides the net phase shift for the reflected wave. Our analysis indicates that not only does the chirality of the massless quasiparticles ensure the absence of back-scattering for normal incidence \cite{klein1, klein2}, but also no splitting under TIR, such as in Ref. \cite{qw1}. Second, we also note the research on the nonlinear screening in the graphene \emph{p-n} junction \cite{nonscreen}, which can lower the junction resistance. Nevertheless, in our scheme [e.g., \fig{fig:pnj}(a)], the shift originates from the evanescent wave penetration into the \emph{p} side having a smaller effective refractive index. The nonlinear screening near the interface can not severely affect the existence of the evanescent wave on the \emph{p} side. In principle, this effect can thus be ignored in our theory.

Based on the above discussions and the stationary phase treatment \cite{nghs1, de1, de2}, the lateral displacement is 
\begin{eqnarray}\label{d}
d&=&\sin \theta \frac{\partial \varphi_{r}}{\partial k_{x}}=-\frac{1}{k_{1}} \frac{\partial \varphi_{r}}{\partial \theta}  \nonumber \\
&=& \frac{ \sin \theta_{c} (s s^\prime \cos^{2} \theta_{c} \sin \theta+\delta (s s ^\prime- \sin \theta_{c}))}{k_{1} \delta (\delta +\sin \theta )\sin \theta(1-s s^\prime \sin \theta_{c})}.
\end{eqnarray}
Note that \eq{d} is not valid at the critical angle $\theta_{c}$ and $\theta = 90\,^{\circ}$. The results depicted in Fig. \ref{fig:d} can be directly analogous to the positive and negative GH shifts in optics \cite{goos, nghs1, nghs2}. Equations (\ref{r}) and (\ref{d}) also tell us that the phase and position of the reflected wave can be coherently controlled by adiabatically changing the gate voltages, which can make the step potential work as a ``movable mirror''.

\begin{figure}[ht]
\centerline{\includegraphics[clip,width=1.0\linewidth]{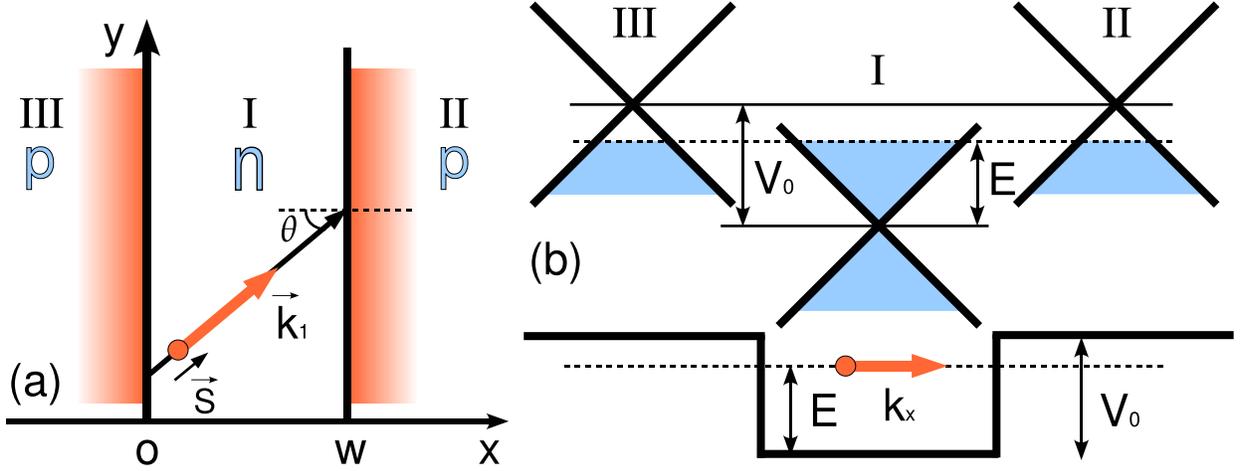}}

\caption{(Color online) (a) Schematic of a symmetric \emph{p-n-p} waveguide with a ray in region $I$ and evanescent fields in regions $II$ and $III$, where the incident angle is $\sin \theta = k_{y}/k_{1} = \hbar v_{f} k_{y}/E$. (b) Low-energy linear spectrum and potential diagram.}

\label{fig:pnp}
\end{figure}

\begin{figure}[ht]
 \includegraphics[clip,width=0.6\linewidth]{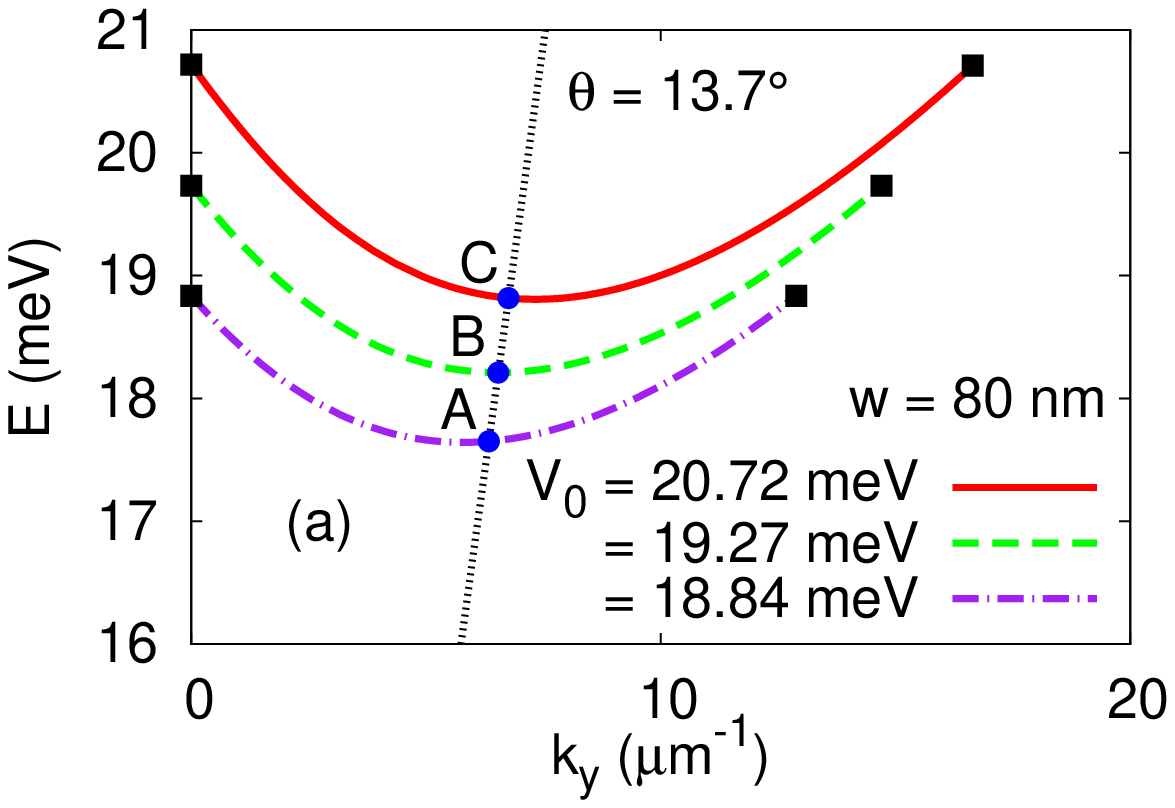}
         \includegraphics[clip,width=0.6\linewidth] {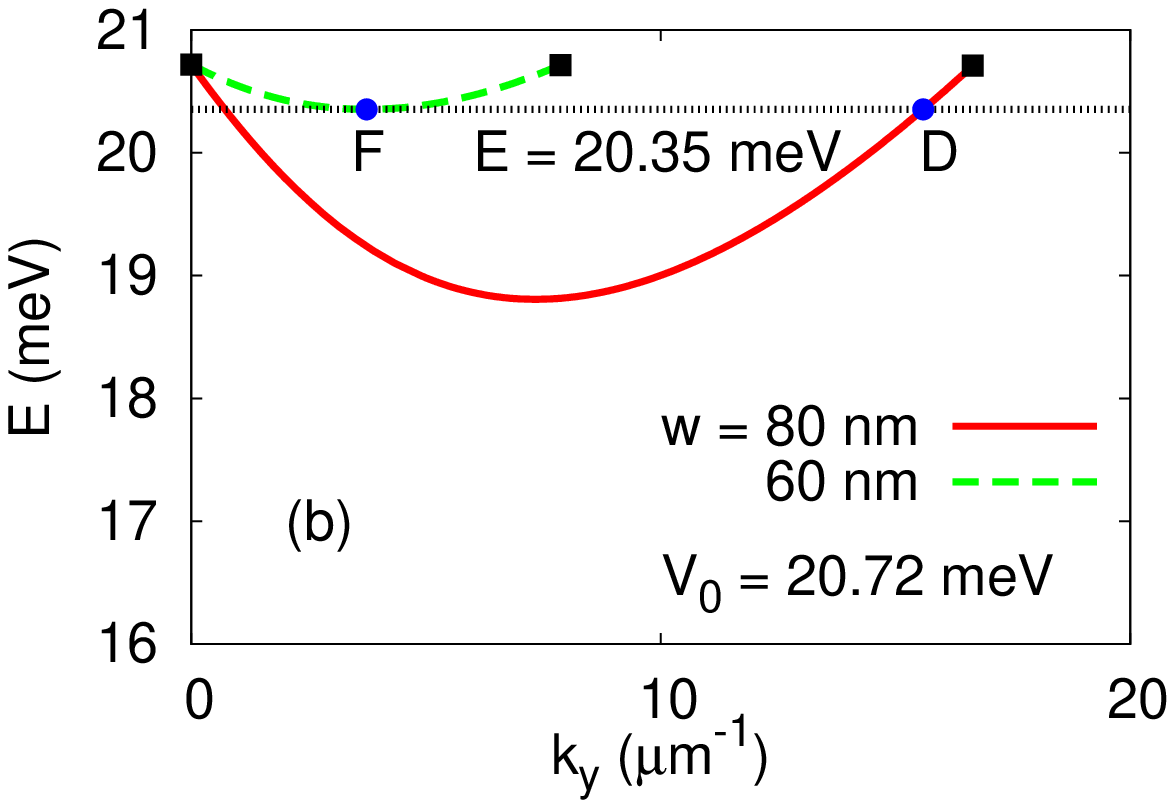}
\caption{(Color online) (a) Dispersion curves for different electrostatic potentials, $V_{0}$ = 18.84 meV (dash-dotted), 19.27 meV (dashed) and 20.72 meV (solid), but the same width of $w=80$ nm. For a particular incident angle $\theta=13.7^{\circ}$ (dotted), one can obtain the different group velocities: $48279.4$ m/s at point A (6.3 $\mu$m$^{-1}$, 17.7 meV), zero at B (6.5 $\mu$m$^{-1}$, 18.2 meV) and -53399.7 m/s at C (6.8 $\mu$m$^{-1}$, 18.8 meV) on the corresponding curves. (b) Dispersion curves for different widths, $w=80$ nm (solid) and $60$ nm (dashed), but the same electrostatic potential of $V_{0}=20.72$ meV. For a particular incident energy $E=20.35$ meV (dotted), one can obtain the different group velocities: $493403.4$ m/s at point D (15.6 $\mu$m$^{-1}$, 20.35 meV) and zero at F (3.7 $\mu$m$^{-1}$, 20.35 meV) on the corresponding curves. Note that our parameters only lead to one single dispersion curve for each pair of $w$ and $V_{0}$, and the solid black squares give the cutoffs for each case.}
   \label{fig:gp}
\end{figure} 

The negative GH-like shift of massless Dirac fermions at the sharp \emph{p-n} junction suggests that, in analogy to the trapped rainbow effect of light in the LHH \cite{tr}, the quasiparticles can have novel propagation properties in a \emph{p-n-p} waveguide. For simplicity, we discuss here a \emph{symmetric} \emph{p-n-p} waveguide [Figs. \ref{fig:pnp}(a) and \ref{fig:pnp}(b)]. Therefore, \eq{ab} is still applicable, but the solutions and boundary conditions are different. In region $III$, we have
\begin{equation}\label{pnps3}
\psi_{III}=A \binom{1}{- i s^\prime (\delta - \sin \theta)/ \sin \theta_{c}}  e^{\alpha x+i k_{y} y} . 
\end{equation}
In region $I$, we have
\begin{eqnarray}\label{pnps1}
\psi_{I}=\binom{B \cos(k_{x} x)+C \sin(k_{x} x)}
{i s (B \sin(k_{x}x + \theta) - C \cos(k_{x}x+\theta))}e^{i k_{y} y}. 
\end{eqnarray}
In region $II$, we have
\begin{eqnarray}\label{pnps2}
\psi_{II}=D \binom{1}{i s^\prime (\delta + \sin \theta) /  \sin \theta_{c}}  e^{-\alpha (x-w)+i k_{y} y},
\end{eqnarray} 
where $w$ is the width of region $I$. The boundary conditions result in
\begin{subequations}\label{bc}
\begin{eqnarray} 
A &=& B,  \\
- A (\delta - \sin \theta)  &=& s s^\prime \sin \theta_{c} (B \sin \theta - C \cos \theta),   \\
D &=& B \cos k_{x} w + C \sin k_{x} w ,   \\
D (\delta + \sin \theta)  &=& s s^\prime \sin \theta_{c} (B \sin (k_{x} w+\theta)  \nonumber  \\
& & - C \cos (k_{x} w + \theta)) . 
\end{eqnarray}
\end{subequations}
A transcendental equation can be derived from the coefficient determinant of \eq{bc}, which is
\begin{eqnarray}\label{te}
\sin \theta_{c} (- s s^\prime \delta \cos \theta  \cos(k_{1} w \cos\theta)+ && \nonumber \\  (\sin \theta_{c}-s s^\prime \sin^{2} \theta) \sin (k_{1} w \cos\theta)) &=& 0. 
\end{eqnarray}
The preceding relationships between $k_{1}$, $k_{2}$, $E$, and $V_{0}$ give the dispersion relations between $E$ and $k_{y}$ for the guided modes. Condition (\ref{ie}) leads to $V_{0}/2<E<V_{0}$, $s=1$, $s^\prime=-1$ for the \emph{p-n-p} waveguide. Numerical calculations of the dispersion relations clearly indicate the slow positive (forward), zero and even negative (backward) group velocities of the quasiparticles depending on the gate voltages [Fig. \ref{fig:gp}(a)] and the widths [Fig. \ref{fig:gp}(b)]. 

More interestingly, the field-effect-dependent dispersion relations shown in Fig. \ref{fig:gp}(a) indicate the possibility to dynamically and coherently decelerate, stop and re-accelerate the guided quasiparticles. The detailed process is as follows. First, it is known that, when the quasiparticles are guided in the waveguide, the incident angle $\theta$ should be preserved due to the parallel interfaces [Fig. \ref{fig:pnp}(a)]. Thus, assuming the quasiparticles are initially prepared in a forward propagating state shown at point A in Fig. \ref{fig:gp}(a), then one can adiabatically changing the gate voltages to transfer the forward propagating state to a trapped state at point B due to the preservation of the incident angle. After a certain storage time, the trapped state can be transferred to either the original state or a backward propagating state shown at point C. Consequently, it can be seen that the \emph{p-n-p} waveguide is not only a waveguide, but also a ``buffer'' or ``memory'' for the quasiparticles in a coherent way.

\begin{figure}[ht]
 \centerline{\includegraphics[angle=0,clip,width=1\linewidth]
{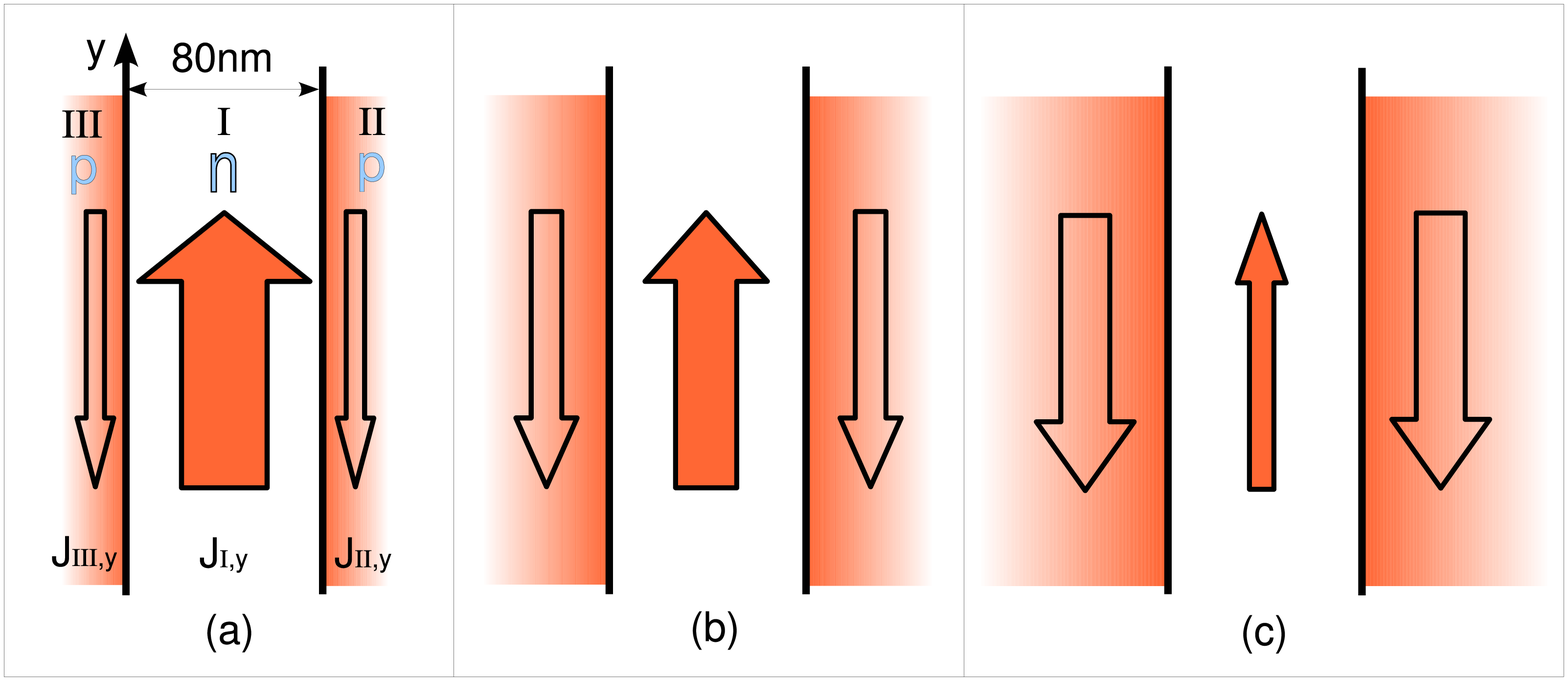}}   \caption{(Color online) Current density diagrams at the points A, B and C in [\fig{fig:gp}(a)], where (a) corresponds to A, (b) to B and (c) to C. The arrows schematically show the magnitudes and directions of the $y$ components of the current densities in the different regions.}
   \label{fig:current}
\end{figure}

To explain this process, we adopt the current-flux method \cite{cfm} instead of the stationary phase treatment mentioned above. The current density $\boldsymbol{J}$ in the different regions can be expressed by  
\begin{equation}\label{j}
\boldsymbol{J}_{n}=v_{f} (\psi_{n}^{\dag} \boldsymbol{\sigma} \psi_{n}) 
\end{equation}
with $n=I,II,III$ for the corresponding regions \cite{qw1}. For the guided modes, we are only interested in their propagation along the \emph{p-n-p} waveguide [the $y$ direction in \fig{fig:pnp}(a)]. Hence, the flux $\Phi$, which also gives the group velocity $v_{g}$ of the quasiparticles, is given by
\begin{equation}\label{jy}
\Phi = v_{g} = \int_{-\infty}^{+\infty} (J_{I,y}+J_{II,y}+J_{III,y}) dx, 
\end{equation}
where $J_{I,y}$ for the guided components gives the positive current density in region $I$, $J_{II,y}$ and $J_{III,y}$ for the evanescent components give the negative current densities in regions $II$ and $III$. By adiabatically changing the gate voltages, we can coherently redistribute the current densities in the different regions. Consequently, the dynamics of the whole wave function are strongly influenced, and the group velocity can be coherently controlled. Our calculations show that \eq{jy} gives the same values of group velocities at the points A, B and C in \fig{fig:gp}(a). The current densities in the \emph{p-n-p} waveguide are schematically illustrated in \fig{fig:current}. 

Additionally, the width-dependent dispersion relations shown in Fig. \ref{fig:gp}(b) also indicate the possibility to coherently decelerate and stop the guided quasiparticles by adiabatically reducing the width from point D to F while keeping the gate voltages unchanged, which is similar to the principle of Ref. \cite{tr}. After the storage, one can adiabatically change the gate voltages for retrieval.

In summary, we have demonstrated the GH-like shifts of massless Dirac fermions at an electrostatic step potential in graphene and the novel dispersion relations for the guided quasiparticle modes in a sharp graphene \emph{p-n-p} waveguide. Our analysis shows that, based on the tunable electric field effects, the electrostatic step potential can function as a movable mirror for the incident quasiparticles, and the \emph{p-n-p} waveguide can coherently buffer, store and retrieve the guided quasiparticles. These results may have interesting applications in the graphene-based quantum information processing. 

We would like to acknowledge funding from NSF. L. Z. thanks A. Kovner, J. Javanainen, R. Zhou, T. Wang and F. Peng for fruitful discussions and helpful suggestions.

\bibliographystyle{apsrev}
\bibliography{trappedrainbow}

\begin{thebibliography}{27}
\expandafter\ifx\csname natexlab\endcsname\relax\def\natexlab#1{#1}\fi
\expandafter\ifx\csname bibnamefont\endcsname\relax
  \def\bibnamefont#1{#1}\fi
\expandafter\ifx\csname bibfnamefont\endcsname\relax
  \def\bibfnamefont#1{#1}\fi
\expandafter\ifx\csname citenamefont\endcsname\relax
  \def\citenamefont#1{#1}\fi
\expandafter\ifx\csname url\endcsname\relax
  \def\url#1{\texttt{#1}}\fi
\expandafter\ifx\csname urlprefix\endcsname\relax\def\urlprefix{URL }\fi
\providecommand{\bibinfo}[2]{#2}
\providecommand{\eprint}[2][]{\url{#2}}

\bibitem[{\citenamefont{Goos and H{\"{a}}nchen}()}]{goos}
\bibinfo{author}{\bibfnamefont{F.}~\bibnamefont{Goos}} \bibnamefont{and}
  \bibinfo{author}{\bibfnamefont{H.}~\bibnamefont{H{\"{a}}nchen}},
  \bibinfo{note}{{Ann. Phys. (Leipzig) {\bf 1}, 333 (1947); {\it ibid.} {\bf
  5}, 251 (1949).}}

\bibitem[{\citenamefont{Artmann}(1948)}]{art}
\bibinfo{author}{\bibfnamefont{K.}~\bibnamefont{Artmann}},
  \bibinfo{journal}{Ann. Phys. (Leipzig)} \textbf{\bibinfo{volume}{2}},
  \bibinfo{pages}{87} (\bibinfo{year}{1948}).

\bibitem[{\citenamefont{Renard}(1964)}]{cfm}
\bibinfo{author}{\bibfnamefont{R.~H.} \bibnamefont{Renard}},
  \bibinfo{journal}{J. Opt. Soc. Am.} \textbf{\bibinfo{volume}{54}},
  \bibinfo{pages}{1190} (\bibinfo{year}{1964}).

\bibitem[{\citenamefont{Berman}(2002)}]{nghs1}
\bibinfo{author}{\bibfnamefont{P.~R.} \bibnamefont{Berman}},
  \bibinfo{journal}{Phys. Rev. E} \textbf{\bibinfo{volume}{66}},
  \bibinfo{pages}{067603} (\bibinfo{year}{2002}).

\bibitem[{ngh()}]{nghs2}
\bibinfo{note}{{I. Shadrivov {\it et al.}, Appl. Phys. Lett. {\bf 83}, 2713
  (2003).}}

\bibitem[{\citenamefont{Snyder and Love}(1983)}]{wg}
\bibinfo{author}{\bibfnamefont{A.~W.} \bibnamefont{Snyder}} \bibnamefont{and}
  \bibinfo{author}{\bibfnamefont{J.~D.} \bibnamefont{Love}},
  \emph{\bibinfo{title}{Optical Waveguide Theory}} (\bibinfo{publisher}{Chapman
  and Hall, New York}, \bibinfo{year}{1983}).

\bibitem[{\citenamefont{Tsakmakidis et~al.}(2007)\citenamefont{Tsakmakidis,
  Boardman, and Hess}}]{tr}
\bibinfo{author}{\bibfnamefont{K.~L.} \bibnamefont{Tsakmakidis}},
  \bibinfo{author}{\bibfnamefont{A.~D.} \bibnamefont{Boardman}},
  \bibnamefont{and} \bibinfo{author}{\bibfnamefont{O.}~\bibnamefont{Hess}},
  \bibinfo{journal}{Nature (London)} \textbf{\bibinfo{volume}{450}},
  \bibinfo{pages}{397} (\bibinfo{year}{2007}).

\bibitem[{g01()}]{g01}
\bibinfo{note}{{K. S. Novoselov {\it et al.}, Science {\bf 306}, 666 (2004).}}

\bibitem[{gre({\natexlab{a}})}]{grev1}
\bibinfo{note}{{A. H. Castro Neto {\it et al.}, arXiv:0709.1163v2.}}

\bibitem[{gre({\natexlab{b}})}]{grev2}
\bibinfo{note}{{C. W. J. Beenakker, Rev. Mod. Phys. {\bf 80}, 1337 (2008).}}

\bibitem[{lcl()}]{lcl}
\bibinfo{note}{{F. Miao {\it et al.}, Science {\bf 317}, 1530 (2007).}}

\bibitem[{pn()}]{pn}
\bibinfo{note}{{K. S. Novoselov {\it et al.}, Nature (London) {\bf 438}, 197
  (2005); Y. Zhang {\it et al.}, Nature (London) {\bf 438}, 201 (2005); H. B.
  Heersche {\it et al.}, Nature (London) {\bf 446}, 56 (2007); M. C. Lemme {\it
  et al.}, IEEE Electron Device Lett. {\bf 28}(4), 282 (2007); B. Huard {\it et
  al.}, Phys. Rev. Lett. {\bf 98}, 236803 (2007); J. R. Williams {\it et al.},
  Science {\bf 317}, 638 (2007)}}.

\bibitem[{kle({\natexlab{a}})}]{klein1}
\bibinfo{note}{{V. V. Cheianov and V. I. Fal'ko, Phys. Rev. B {\bf 74},
  041403(R) (2006).}}

\bibitem[{kle({\natexlab{b}})}]{klein2}
\bibinfo{note}{{M. I. Katsnelson {\it et al.}, Nature Phys. {\bf 2}, 620
  (2006).}}

\bibitem[{nr1()}]{nr1}
\bibinfo{note}{{V. V. Cheianov {\it et al.}, Science {\bf 315}, 1252 (2007).}}

\bibitem[{qw1()}]{qw1}
\bibinfo{note}{{J. Milton Pereira, Jr. {\it et al.}, Phys. Rev. B {\bf 74},
  045424 (2006).}}

\bibitem[{qw2()}]{qw2}
\bibinfo{note}{{J. Milton Pereira, Jr. {\it et al.}, Appl. Phys. Lett. {\bf
  90}, 132122 (2007).}}

\bibitem[{gst()}]{gstrip1}
\bibinfo{note}{{J. Tworzyd{\l}o, {\it et al.}, Phys. Rev. Lett. {\bf 96},
  246802 (2006).}}

\bibitem[{\citenamefont{Katsnelson}(2006)}]{gstrip2}
\bibinfo{author}{\bibfnamefont{M.~I.} \bibnamefont{Katsnelson}},
  \bibinfo{journal}{Eur. Phys. J. B.} \textbf{\bibinfo{volume}{51}},
  \bibinfo{pages}{157} (\bibinfo{year}{2006}).

\bibitem[{\citenamefont{Silvestrov and Efetov}(2007)}]{qd1}
\bibinfo{author}{\bibfnamefont{P.~G.} \bibnamefont{Silvestrov}}
  \bibnamefont{and} \bibinfo{author}{\bibfnamefont{K.~B.}
  \bibnamefont{Efetov}}, \bibinfo{journal}{Phys. Rev. Lett.}
  \textbf{\bibinfo{volume}{98}}, \bibinfo{pages}{016802}
  (\bibinfo{year}{2007}).

\bibitem[{spi()}]{spin}
\bibinfo{note}{{B. Trauzettel, {\it et al.}, Nature Phys. {\bf 3}, 192
  (2007).}}

\bibitem[{qd2()}]{qd2}
\bibinfo{note}{{ A. Matulis {\it et al.}, Phys. Rev. B {\bf 77}, 115423
  (2008).}}

\bibitem[{bf()}]{bf}
\bibinfo{note}{{A. De Martino {\it et al.}, Phys. Rev. Lett. {\bf 98}, 066802
  (2007).}}

\bibitem[{in()}]{in}
\bibinfo{note}{{ R. Rold{\'{a}}n {\it et al.}, Phys. Rev. B {\bf 77}, 115401
  (2008).}}

\bibitem[{\citenamefont{Fradkin and Kashuba}()}]{de1}
\bibinfo{author}{\bibfnamefont{D.~M.} \bibnamefont{Fradkin}} \bibnamefont{and}
  \bibinfo{author}{\bibfnamefont{R.~J.} \bibnamefont{Kashuba}},
  \bibinfo{note}{{Phys. Rev. D {\bf 9}, 2775 (1974); {\it ibid.} {\bf 10}, 1137
  (1974).}}

\bibitem[{de2()}]{de2}
\bibinfo{note}{{S. C. Miller, Jr. {\it et al.}, Phys. Rev. Lett. {\bf 29}, 740
  (1972).}}

\bibitem[{non()}]{nonscreen}
\bibinfo{note}{{L. M. Zhang, {\it et al.}, Phys. Rev. Lett. {\bf 100}, 116804
  (2008).}}

\end{thebibliography}

\end{document}